# Colloidal Manganese Doped ZnS Nanoplatelets and Their Optical Properties

Liwei Dai,[a] Christian Strelow,[a] Tobias Kipp,[a] Alf Mews,[*,a] Iris Benkenstein,[b] Dirk Eifler,[b] Thanh Huyen Vuong,[c] Jabor Rabeah,[c] James McGettrick,[d] Rostyslav Lesyuk,[e,f] Christian Klinke[*,f,g]

[a]*Institute of Physical Chemistry, University of Hamburg, Martin-Luther-King-Platz 6, 20146 Hamburg, Germany*
[b]*Department of Chemistry, University of Hamburg, Martin-Luther-King-Platz 6, 20146 Hamburg, Germany*
[c]*Leibniz Institute for Catalysis, Albert-Einstein-Straße 29a, 18059 Rostock, Germany*
[d]*SPECIFIC-IKC, Materials Research Centre, College of Engineering, Swansea University, Bay Campus, Swansea SA1 8EN, UK*
[e]*Pidstryhach Institute for Applied Problems of Mechanics and Mathematics of NAS of Ukraine, Naukowa str. 3b, 79060 Lviv & Department of Photonics, Lviv Polytechnic National University, Bandery str. 12, 79000 Lviv, Ukraine*
[f]*Institute of Physics, University of Rostock, Albert-Einstein-Straße 23, 18059 Rostock, Germany*
[g]*Department of Chemistry, Swansea University – Singleton Park, Swansea SA2 8PP, UK*

**Abstract:** Manganese (Mn)-doped ZnS nanocrystals (NCs) have been extensively explored for optical applications with the advantages of low toxicity, large Stokes shifts and enhanced thermal and environmental stability. Although numerous studies on Mn-doped ZnS dots, rods and wires have been reported, the literature related to Mn-doped ZnS nanoplatelets (ZnS:Mn NPLs) is scarce. Here, we present the first example of direct doping of $Mn^{2+}$ ions into ZnS NPLs via the nucleation-doping strategy. The resulting ZnS:Mn NPLs exhibit Mn luminescence, indicative for successful doping of the host ZnS NPLs with $Mn^{2+}$ ions. The energy transfer from the ZnS NPLs to the $Mn^{2+}$ ions was observed by employing spectroscopic methods. Furthermore, the impact of the Mn concentration on the optical properties of ZnS:Mn NPLs was systematically investigated. As a result of Mn-Mn interaction, tuneable Mn emission and shortened photoluminescence (PL) lifetime decay were observed and rationalized by means of electron paramagnetic resonance (EPR) and X-ray photoelectron spectroscopy (XPS). Finally, we show that the initially low dopant-PL quantum yield (QY) of ZnS:Mn NPLs can be dramatically enhanced by passivating the surface trap states of the samples. The presented synthetic strategy of ZnS:Mn NPLs opens a new way to synthesize further doped systems of two-dimensional (2D) NPLs.

**INTRODUCTION**

Quasi two-dimensional (2D) nanoplatelets (NPLs) with atomically precise thicknesses have emerged as a novel class of materials.[1-4] Such 2D NPLs are appealing due to their large surface-to-volume ratio, strong quantum confinement effects, and giant oscillator strength, making them promising candidates for applications in lasing, energy conversion and storage, and optoelectronic devices.[5-7] In contrast to zero-dimensional (0D) and one-dimensional (1D) nanocrystals (NCs), 2D NPLs present some unique optoelectronic properties including extremely narrow spectral line widths in absorption and emission and relatively large absorption cross sections because of their well-defined thickness.[8] However, these NPLs also display some shortcomings originating from their intrinsic structural limitations. For instance, the optical properties of NPLs lack continuous tunability since the thickness variation of NPLs is discrete.[9] Additionally, the emission spectrum of NPLs is limited to a relatively narrow wavelength range due to the finite number of atomic planes in the thickness direction.[10] A typical example are CdSe NPLs. Although CdSe NPLs have been synthesized with variable thicknesses, their emission spectrum can only be tuned in a range from 460 to 625 nm.[1, 11] Similarly, the ZnS NPLs we reported previously exhibit spectrally fixed photoluminescence (PL) due to the lack of thickness tunability.[12]

Doping of NCs, as an effective strategy to modify their optical properties, has attracted intensive research interests in the past years. With a small amount of dopants incorporated into the NCs, they can exhibit many new optical, electronic and magnetic phenomena.[13-14] Recently, Mn-doped ZnS NCs with low toxicity have been explored. After the first synthesis of Mn-doped ZnS quantum dots (QDs),[15] a variety of synthetic methods to prepare high-quality Mn-doped ZnS QDs have been developed.[16-18] In addition to the doped QDs, Liu et al. doped ZnS quantum rods (QRs) with $Mn^{2+}$ ions which displayed high quantum yield (QY) (up to 45%).[19] Despite a lot of studies report Mn-doped ZnS dots, rods and wires, the literature related to Mn-doped ZnS (ZnS:Mn) NPLs is scarce. Zinc sulfide NPLs present an attractive choice as the matrix for $Mn^{2+}$ ions. Zinc and manganese ions in tetrahedral coordination have similar ionic radii (74 vs 80 pm).[20-21] Due to the strong confinement in NPLs, the matrix absorption and $Mn^{2+}$ emission can be spaced in spectrum by more than 2 eV. Additionally, the wurtzite (WZ) phase of ZnS shows enhanced crystal field and spontaneous polarization being able potentially to influence the PL of the dopant.[22] Altogether, this creates a perspective non-explored platform for low-toxic energy conversion units with no self-absorption and enhanced oscillator strength. To date, one work showed the preparation of ZnS:Mn NPLs by the post-synthesis treatment,[23] whereas the as-synthesized ZnS:Mn NPLs were quite unstable. The $Mn^{2+}$ ions are very easily ejected from the ZnS NPLs once the temperature is below 180 °C. This research reflects that the synthesis of ZnS:Mn NPLs still remains a big challenge.

Herein, we report a simple, effective and "green" (phosphine-free) synthesis method for obtaining ZnS:Mn NPLs with uniform thickness. The as-synthesized ZnS:Mn NPLs display Mn luminescence indicating the successful doping of the ZnS crystal lattice with $Mn^{2+}$ ions. To the best of our knowledge, this is the first time of direct doping of colloidal ZnS NPLs with $Mn^{2+}$ ions via a nucleation-doping strategy. We explore the mechanism of energy transfer between ZnS NPLs and $Mn^{2+}$ ions by using steady-state UV-Vis absorbance, PL and PL excitation (PLE) measurements. The effect of different Mn concentrations on the optical properties of ZnS:Mn NPLs is investigated in detail. Steady-state PL spectra, PL QY measurements and time-resolved PL decay of the samples reveal red-shifted Mn emission, relatively low PL QY and short PL lifetimes due to the presence of strong Mn-Mn interaction. Furthermore, we show that the Mn-related PL QY of doped NPLs can be dramatically enhanced by passivating the surface trap states of the samples.

**RESULTS AND DISCUSSION**

In this study, colloidal Mn-doped ZnS NPLs were synthesized by adapting the soft-template method for the synthesis of ZnS NPLs, which we reported before.[12] The original method was already optimized for obtaining the most stable and homogeneous product. For this, the parametric window was found to be very narrow. It turns out that all the time, temperature, and species-concentration regimes of the ZnS NPL synthesis also work best for the synthesis of Mn-doped ZnS NPLs. In a typical synthesis, zinc chloride, manganese acetate, and sulfur powder with a nominal Mn:Zn:S molar ratio of $x$:1:3 were dissolved in a mixture of oleylamine (OAm) and octylamine (OTA). In this study, $x$ has been varied between approximately 0.03% and 16%. After purging with nitrogen at 100 °C for half an hour, the reaction solution was heated to 150 °C for 6 hours. After cool-down, precipitation and washing of the samples, a transmission electron microscopy (TEM) grid was prepared. The exemplary TEM image in Figure 1A of the as-synthesized Mn-doped ($x$ = 4%) ZnS nanostructures reveal their platelet-like shape similar to undoped NPLs, as shown in Figure 1B. A high-resolution (HR) TEM image of an individual ZnS:Mn NPL (Figure 1C) exhibits the well-resolved lattice fringe pattern illustrating that the doped NPLs are well-crystallized. The observed lattice spacings were measured to be 3.09 and 3.27 Å, matching well the (0001) and (10-10) plane spacings of the bulk WZ-ZnS structure (ICPDS 00−080−0007). The corresponding fast Fourier transform (FFT) (Figure 1D) clearly reveals that the [0001] and [10-10] directions span the basal plane of the NPL. The X-ray diffraction (XRD) pattern for the ZnS:Mn NPLs (Figure 1E) indicates that they retain the WZ structure of ZnS (ICPDS 00−080−0007) after Mn doping. The energy-dispersive X-ray spectroscopy (EDS) spectrum (Figure S1, Supporting Information) confirms the presence of an Mn component in NPLs suggesting the atomic ratio of Mn:Zn to be 0.02:1, different to the ratio of 0.04:1 as set for the precursors. To accurately determine the content of Mn in NPLs, inductively coupled plasma optical emission spectroscopy (ICP-OES) was performed and will be discussed below. Additionally, the EDS spectrum in Figure S1 shows the atomic ratio of Zn:S to be 0.91:1, which is close to the stoichiometric ratio of ZnS indicating slight zinc deficiency.

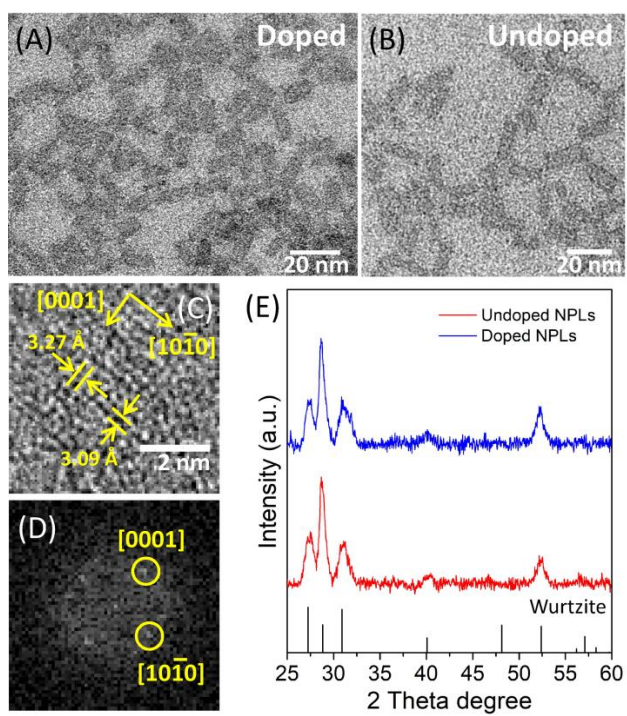

**Figure 1.** TEM images of doped (A) and undoped (B) NPLs synthesized with an Mn:Zn:S ratio of 0.04:1:3 and 0:1:3 in the synthesis, respectively. (C) HRTEM image of an individual doped NPLs and (D) the corresponding FFT pattern. (E) XRD patterns of undoped NPLs (red) and doped NPLs (blue). Vertical lines indicate the bulk wurtzite ZnS pattern (ICDD card no. 00−080−0007).

The optical properties of doped NPLs ($x = 4\%$) were characterized by steady-state UV−Vis absorbance and PL spectroscopy. The absorbance spectra of doped and undoped samples (Figure 2A, left) show nearly identical spectral features, manifesting that the electronic structure of the host remains unchanged after doping. The narrow excitonic absorption peak at a wavelength of 283 nm (4.38 eV) reflects a high uniformity in thickness (5 monolayer thick, 1.8 nm) of the doped NPLs.[12] The PL spectra of the doped ($x = 4\%$) and undoped NPLs shown in Figure 2A (right) have been recorded for an excitation wavelength of 260 nm. The prominent peak in the spectra at 520 nm is not a PL signal but a signal originating from the second diffraction order of the excitation light. A smaller peak with a wavelength of approximately 560 nm can be assigned to the second diffraction order of the Raman signal from the solvent (hexane). The PL spectrum of the doped NPLs exhibits an emission band centered at 601 nm (2.06 eV) with a full width at half-maximum (FWHM) of about 237 meV, which is not observed in the PL spectrum of undoped NPLs. This band represents the typical orange emission of Mn-doped ZnS samples, which can be observed by eye under UV illumination, as proven by a photograph of a NPL-coated cover slide shown in the inset of Figure 2A. It can be assigned to the $Mn^{2+}$ ion $^4T_1$-$^6A_1$ transitions,[16, 24] and it is a clear indication of the successful doping of the ZnS crystal lattice with $Mn^{2+}$ ions. Figure 2B shows a PLE spectrum of doped NPLs (green data points) for which the detection wavelength was set to 600 nm. The intense feature at 300 nm is again a signal due to the second diffraction order of the spectrometer grating. The actual PLE spectrum resembles the absorption spectrum (dashed line), revealing that after ZnS excitation an energy transfer from the host material to the dopant $Mn^{2+}$ ions occurs before the dopant-PL is emitted. The finding of the energy transfer between the host and the dopant is supported by comparing the PL spectra of undoped and doped NPLs, as shown in Figure 1C. Here, an excitation wavelength of 280 nm was used for two purposes. First, this excitation wavelength is in resonance with the excitonic absorption, which significantly increases the PL intensity of the sample (see comparison in Figure S2 in the Supporting Information). Second, the spectral overlap between the Raman peak of the solvent (hexane) and the exciton emission of the ZnS NPLs is minimized. Without Mn doping, the NPLs have a discernable emission band centered at 292 nm (see black curve in Figure 2C), which originates from the exciton recombination of ZnS NPLs.[12] With Mn doping, the excitonic PL is nearly quenched (see red curve in Figure 2C) suggesting a strong coupling of excitons and Mn dopants in NPLs. The above discussion with respect to the energy transfer between the host and dopant is summarized in the energy scheme in Figure 2D.

In order to investigate the effects of the number of $Mn^{2+}$ ions per NPL on the optical properties and the energy transfer, a set of samples of doped NPLs were synthesized for which the nominal Mn:Zn atomic ratio of the precursors was tuned from approximately 0.03% to 16% while all other experimental parameters were the same. The XRD patterns of these samples (see Figure S3 in the Supporting Information) show no significant changes with varying the doping level, implying that the crystal structure of doped NPLs is almost unaltered even at high doping levels. The actual Mn:Zn atomic ratio in the doped NPLs was determined by ICP-OES. Table 1 compares the Mn:Zn ratios used in the syntheses and measured in the NPL samples. Obviously, the actual Mn:Zn molar ratio in the NPLs is generally lower than the nominal ratio in the reaction solution, indicating that $Mn^{2+}$ ions are less reactive during the synthesis than $Zn^{2+}$ ions. A correlation between both molar ratios (see Figure S4) reveals that the strongest deviations occur for medium doping levels, while for very low (0.03%) and very high (16%) doping levels the ICP-OES-determined Mn:Zn ratio is essentially the same as in the corresponding synthesis solution. If we assume average NPL dimensions of 15.8 nm and 6.4 nm in length and width (as measured from TEM images) and 1.8 nm in height, the average number of $Mn^{2+}$ ions per NPL can be estimated. For our set of samples, the average number of Mn ions per NPL ranges between 1.4 and 793, as summarized in Table 1.

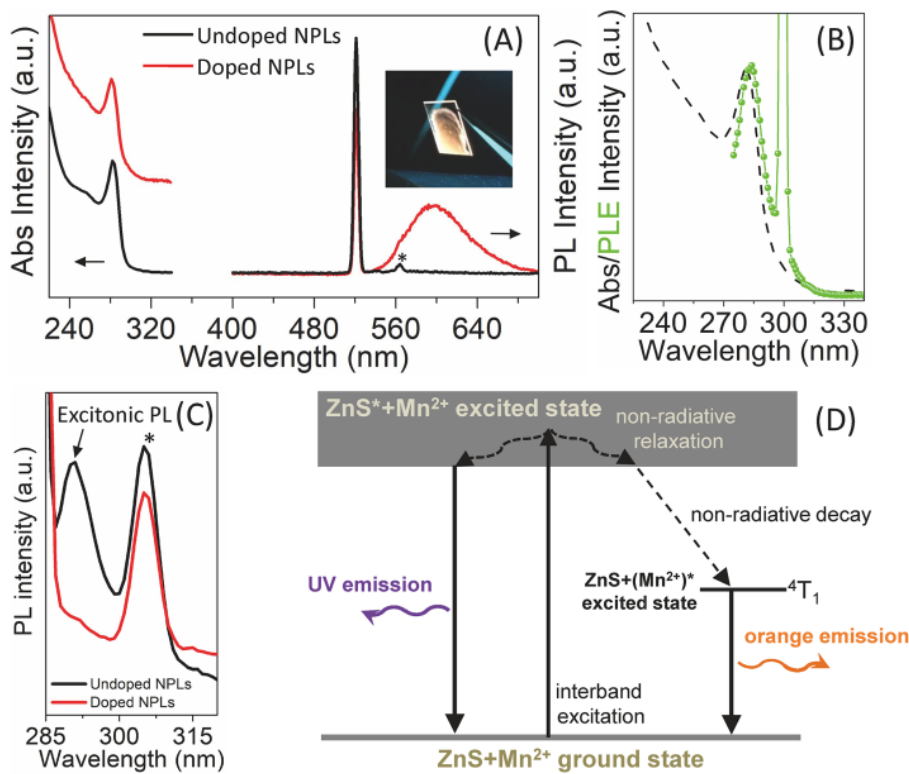

**Figure 2.** (A) Left: UV-Vis absorbance spectra of undoped (black) and doped (red) NPLs. The spectra are vertically shifted for clarity. Right: Corresponding room-temperature PL emission spectra ($\lambda_{exc}$ = 260 nm). The inset shows a photography of the doped NPL dispersion under UV illumination. (B) PLE spectrum of doped NPLs monitored at 600 nm. (C) Room-temperature PL emission spectra of undoped (black) and doped (red) NPLs ($\lambda_{exc}$ = 280 nm). (D) Schematic representation of the photophysical processes in ZnS:Mn NPLs. The scattered light of the lamp (narrow peaks at 520 nm in Figure 2A and 300 nm in Figure 2B) and the solvent (hexane) Raman peaks (marked with *) are also detected.

**Table 1.** Left: nominal Mn:Zn atomic ratios used in the synthesis. Middle: the actual Mn:Zn atomic ratios in ZnS:Mn NPL samples determined by ICP-OES. Right: the average number of Mn ions incorporated per ZnS:Mn NPL determined via ICP-OES.

| Nominal Mn:Zn atomic ratio (by precursor) | ICP-OES results of Mn:Zn atomic ratio | Average number of Mn ions per NPL |
|---|---|---|
| 0.03% | 0.03% | 1.4 |
| 0.06% | 0.04% | 1.9 |
| 0.13% | 0.06% | 2.9 |
| 0.25% | 0.10% | 4.9 |
| 0.50% | 0.20% | 9.5 |
| 1% | 0.38% | 18 |
| 2% | 0.69% | 33 |
| 4% | 1.52% | 73 |
| 8% | 4.89% | 233 |
| 16% | 16.64% | 793 |

Figure 3A (left) shows UV-Vis absorbance spectra for the set of samples of different doping levels. As the doping level increases from approx. 0.03% to 1.52% Mn:Zn atomic ratio in the NPLs (corresponding to up to 73 Mn ions per NPL), no other particular features occur for the ZnS:Mn NPLs as compared to the undoped NPLs. For even stronger doping, the excitonic absorption peak broadens slightly (Figure 3B), which could be a result of the $Mn^{2+}$ ion influence on the electronic structure of host NPLs.[25] The dopant-PL spectra of ZnS:Mn NPLs, also shown in Figure 3A (right), exhibit a broad peak with a nearly stable value for the FWHM of 0.24-0.29 eV. The dopant-PL peak position, however, clearly red-shifts with increasing doping level, as can be seen from the graph in Figure 3C. This red-shift can be attributed to the strong Mn-Mn interaction.[26-27]

The relationship between the number of $Mn^{2+}$ ions per NPL and the energy transfer was investigated by monitoring the exciton emission of the NPLs. To directly compare the intensity of the exciton emission, the concentration of the samples was maintained consistent (the optical density of the samples at the maximum of the excitonic peak is fixed at a certain value). Figure S5A shows that the PL spectrum of undoped NPLs exhibits an obvious exciton emission peak. With increasing number of $Mn^{2+}$ ions, the intensity of the peak drops and the peak nearly disappears when each NPL is doped with ca. 73 Mn atoms on average. This result indicates that the evident energy transfer can occur once the certain concentration of luminescent centers is reached. Additionally, it was found (Figure S5B) that the intensity ratio of exciton emission and Mn emission decreases from 0.27 to 0 as the Mn doping level was increased from 0.03% to 16.64%.

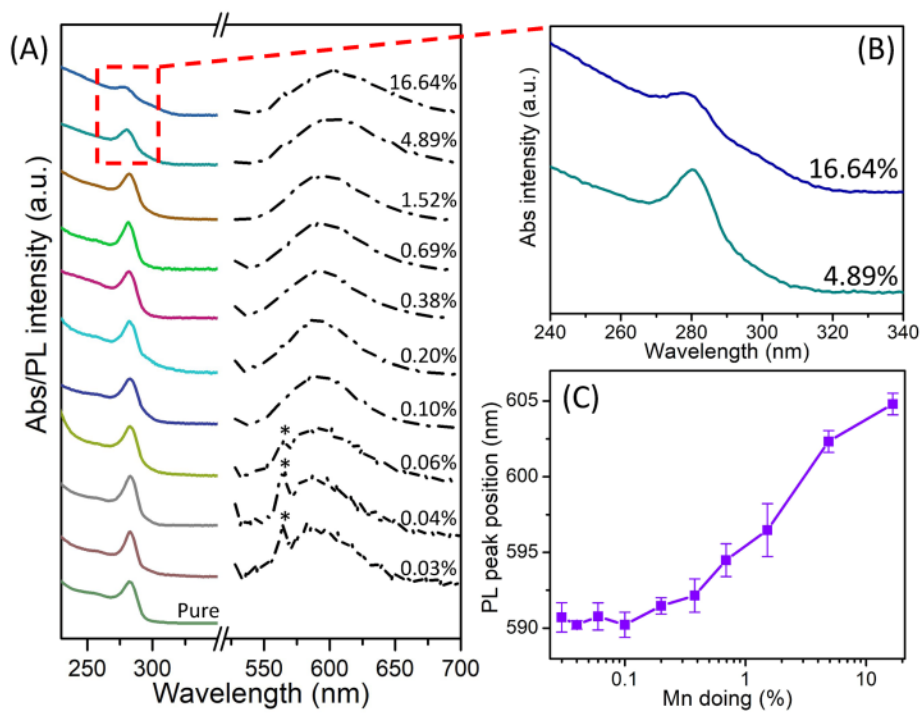

**Figure 3.** (A) Steady-state UV-Vis absorbance and PL spectra of NPLs with different Mn doping levels determined by ICP-OES ($\lambda_{exc}$ = 260 nm). The solvent (hexane) Raman peaks (marked with *) also show up. (B) A magnified view of the UV-Vis absorbance spectra for doped NPLs. (C) Plot of the dopant-PL peak position against the Mn doping level in the synthesis.

The exact location of $Mn^{2+}$ ions in NPLs was explored by using electron paramagnetic resonance (EPR) spectroscopy. At a low doping level (0.04%), the EPR spectrum (Figure 4, black) of the sample exhibits multiplet hyperfine splitting lines, indicating the existence of several Mn species. Two weak six-line signals of Mn(II) at g = 2.004 with resolved hyperfine structure (A = 70 G and 92 G, zero-fine-splitting parameter D = 410 MHz and 189 MHz, respectively) can be extracted from the EPR spectrum of the sample (0.04%), indicating the presence of different Mn species. The one with high degree of symmetry (A = 70 G) corresponds to the tetrahedral symmetry of $Mn^{2+}$ incorporated into the ZnS lattice[16, 25, 28] and the other one with low degree of symmetry (A = 92 G) is close to surface/edges or on the surface of NCs,[29-30] respectively. The observation of different Mn species indicates that $Mn^{2+}$ ions were not only doped inside the NPLs, but also on the surface of NPLs during the doping process. As the doping level increases, the contribution of hyperfine peaks gradually decreased and nearly disappeared at a doping level of 4.89% (Figure 4, blue), whereas a broad signal developed and became more and more apparent. This broad signal originating from the Mn-Mn interaction indicates strong magnetically interacting $Mn^{2+}$ species and/or the formation of Mn aggregations.[31]

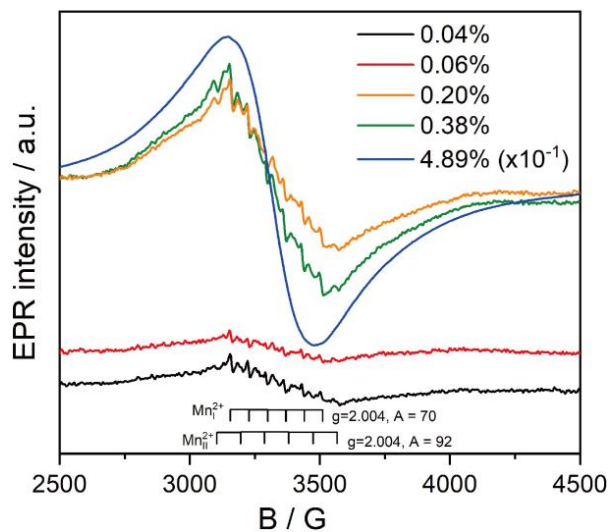

**Figure 4.** EPR spectra of ZnS:Mn NPLs with different Mn doping levels determined by ICP-OES. Two sets of vertical lines indicate the sextet lines hyperfine splitting of 70 and 92 G, respectively.

To study the chemical state of the samples, X-ray photoelectron spectroscopy (XPS) was performed. The XPS survey spectrum in Figure S6 shows the presence of Zn, S, Mn, O, C, and N elements in the doped NPL samples. The existence of C and N elements is due to the OAm and OTA capping the surface of the NPLs. The Zn-2p core level XPS spectrum (Figure 5A) splits into Zn-2p$_{3/2}$ (1021.7 eV) and Zn-2p$_{1/2}$ (1044.8 eV) with a spin-orbit splitting energy of 23.1 eV, which is consistent with that of Zn$^{2+}$.[32] The deconvolution of the S-2p core level spectrum (Figure 5B) shows two peaks centered at 161.8 eV (S-2p$_{3/2}$) and 163 eV (S-2p$_{1/2}$), confirming the presence of the bivalent S$^{2-}$ state with its characteristic peak separation of 1.2 eV. The S-2p$_{3/2}$ signal centered at 161.8 eV is attributed to Zn-S bonds.[33] The asymmetric Mn-2p$_{3/2}$ core level spectrum (Figure 5C) was fitted with the models used by Biesinger et al.[34] The multiplet peaks centered in the region 640-645 eV are assigned to Mn (II),[34] suggesting that Mn$^{2+}$ ions are placed in slightly different electronic environment. This different local environment of Mn$^{2+}$ ions could be caused by their different degrees of coupling. Additionally, a small shake-up peak around 646 eV can be assigned to surface oxidation of Mn (Mn-O),[34] which could be a sign of the presence of surface Mn. The O 1s core level spectrum (Figure 5D) is deconvoluted into three peaks located at 530.8, 531.9, and 533.3 eV, respectively. The former corresponds to oxygen in metal oxides (O$^{2-}$),[35] and might be partially attributed to Mn$^{2+}$ conjugated to O$^{2-}$. The latter two components are due to O$^{2-}$ in the oxygen-deficient regions and the surface hydroxyl groups (O−H) adsorbed on the sample surfaces.[36] As mentioned above, the EPR results show the presence of several Mn species: some incorporated into the ZnS lattice (inner Mn) and the other ones located presumably on the surface of NPLs (surface Mn), which is further supported by the XPS spectrum analysis.

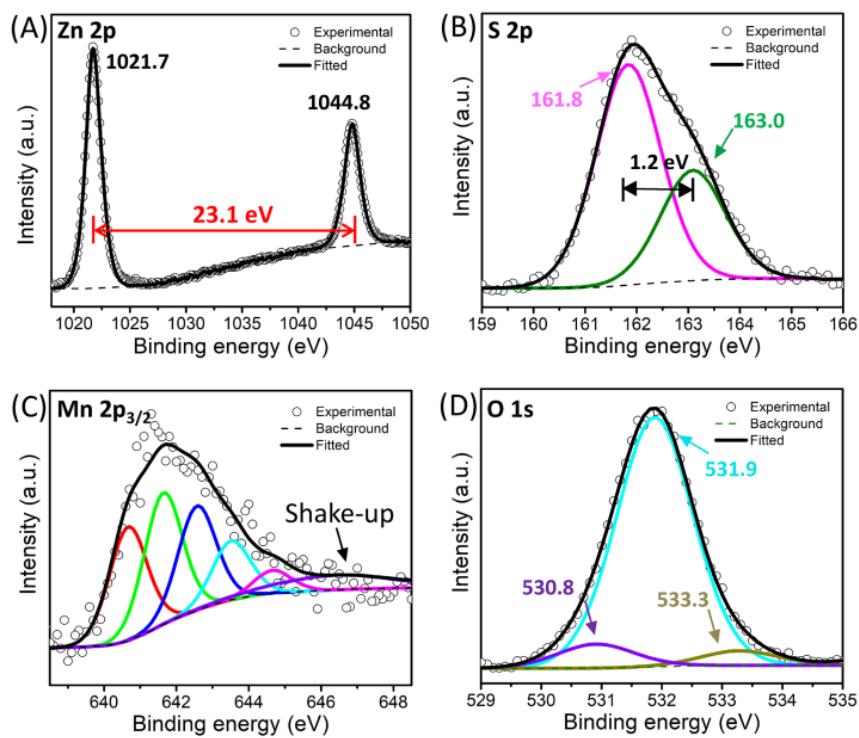

**Figure 5.** Core-level XPS spectra of (A) Zn 2p, (B) S2p, (C) Mn $2p_{3/2}$, and (D) O 1s regions of ZnS:Mn NPLs synthesized with an Mn:Zn:S ratio of 0.04:1:3 in the synthesis.

From the XPS spectra, the ratio of Mn:Zn can be found (Table S1). If we compare the nominal Mn:Zn ratio, the ratio found by IPC-OES and the one by XPS we note that the content of Mn measured by XPS is much higher than that measured by ICP-OES in all samples and exceeds the nominal values given by the synthesis. We attribute the discrepancy to two factors: first, despite the ultrathin structure of NPLs the surface sensitivity of XPS might still contribute.[9] Since the ICP-OES probes the entire composition of the material by dissolving the NPLs. XPS would then suggest that $Mn^{2+}$ might occupy the surface more readily than the volume. The second factor is the slight zinc-deficiency of NPLs observed in EDS, which is attributed to the zinc surface vacancies. Zinc treatment, which will be discussed in the following section, confirms that. Hence, even at a low doping level (0.2%), the XPS ratio still displays much higher values of Mn atoms comparable to the nominal values, revealing that Mn atoms readily occupy the surface of NPLs during the doping process. This phenomenon can be well explained by the "self-purification" effect[37-38] usually observed in ultra-small or ultra-thin nanostructures. Due to the small interior volume of the ultrathin NPLs, the $Mn^{2+}$ ions tend to be expelled and easily migrate to the surface. Along with close ionic radii of $Zn^{2+}$ and $Mn^{2+}$, "self-purification effect" is responsible for the invariability of the XRD spectra of NPLs with high doping levels up to 16%. Additionally, the geometrical peculiarity of the ultrathin NPLs offers the following advantage: dopants (such as Mn ions) induce less stress due to the smaller packing density in comparison to spherical particles. For idealized NPLs with dimensions of 20 nm × 20 nm × 1.8 nm and random dopant distribution, the average distance between dopant ions is estimated to be 1.6 times larger than of the quantum dot with identical volume and same number of dopants (Monte-Carlo simulations, see Figure S7 and Figure S8) that also can contributes to the invariability of the XRD spectra.

To further investigate the dopant-PL properties of the samples, time-resolved PL measurements were performed. Figure S9 shows exemplary PL decay curves for different doping levels. All curves exhibit a non-single-exponential behavior, suggesting an inhomogeneous local environment of $Mn^{2+}$ ions in NPLs, which is in agreement with the EPR and XPS results. The PL decay curves could be fitted with a biexponential function and a time constant at 1/e of the PL intensity is defined as the PL decay time. At a low doping level (0.04%), the time constant of the PL decay is 0.61 ms, which is shorter than previously reported for Mn-doped ZnS NCs (typically >1 ms).[19, 39] This shortened Mn emission lifetime can be assigned to the magnetically coupled $Mn^{2+}$ ions proven by the previous EPR experiments. Generally, the overall PL decay becomes faster with increasing doping level; for a doping level of 16.64%, a time constant of 0.05 ms was found. This shortening can be assigned to an increased Mn-Mn interaction.

Figure 6A demonstrates the quantum yield (QY) of the dopant-PL emission of ZnS:Mn NPLs as function of the Mn-doping level. It is found that the dopant-PL QY increases with increasing doping level and reaches a maximum QY of 1.25% at a doping level of 0.69%. This initial increase can be attributed to the presence of more $Mn^{2+}$ luminescent centers. With further increase in the doping level, the dopant PL QY decreases gradually in particular due to the concentration quenching effect.[40] From the QY results, the doping level of 0.69% seems to be a critical point where the Mn-Mn interaction becomes dominant, which strongly influences the optical properties of doped NPLs. It is also possible to evaluate the QY contribution per $Mn^{2+}$ ion for the different samples, as shown in Figure S10. We find that the QY contribution per $Mn^{2+}$ ion decreases monotonically over the whole doping range. This is probably due to competitive effects of introduction of more luminescent centers, self-purification, and introduction of more strain and defects on the surface and inside.

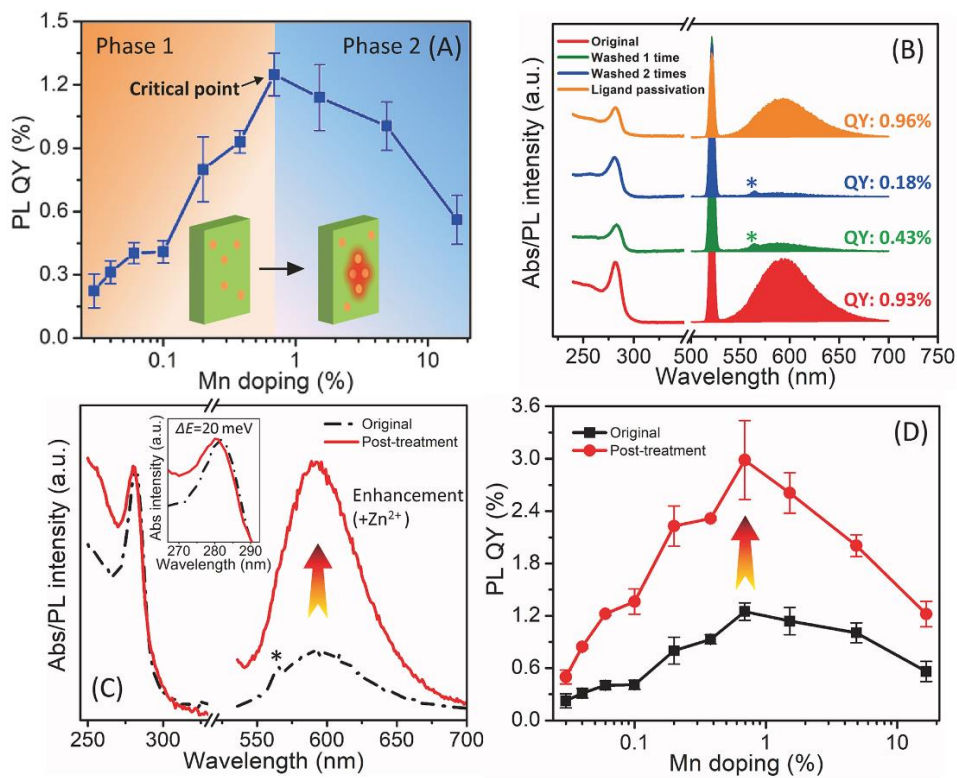

**Figure 6.** (A) Plot of the dopant-PL QY of ZnS:Mn NPLs as a function of the doping level. (B) Left: UV-Vis absorbance spectra of ZnS:Mn NPLs synthesized with an Mn:Zn:S ratio of 0.01:1:3 in the synthesis. Right: Corresponding room-temperature PL emission spectra ($\lambda_{exc}$ = 260 nm). (C) Left: UV-Vis absorbance spectra of ZnS:Mn NPLs before (black) and after (red) post-treatment with a $Zn^{2+}$ solution. The inset shows a slight blue-shift in the absorption peak of ZnS:Mn NPLs. Right: Corresponding room-temperature PL emission spectra ($\lambda_{exc}$ = 260 nm) of the sample. (D) Plot of the dopant PL QY of ZnS:Mn NPLs as a function of the doping level. The 2nd order scattering of the solvent (hexane) Raman peaks (marked with *) and the lamp peaks (narrow peaks at 520 nm) are also visible.

The generally low dopant-PL QY (0.22-1.25%) of ZnS:Mn NPLs is probably related to surface trap states due to the high surface-to-volume ratio. To investigate the influence of the surface trap states, we performed the following control experiments exemplarily for the ZnS:Mn NPL sample with estimated 18 Mn ions per NPL (0.38% doping level by ICP-OES). Firstly, the dopant-PL QY of the original OAm/OTA-capped sample was measured to be 0.93% (Figure 6B, red). After washing once, the QY of the sample dropped significantly to 0.43% (cf. Figure 6B, green). A further decrease in the QY of the sample was observed after a second washing step (cf. Figure 6B, blue). According to previous reports,[41-42] the decreased PL QY upon removal of ligands could be attributed to two mechanisms: (1) The formation of unstable structures and (2) the exposure of more surface trap states. The first possibility can be ruled out because the UV-Vis absorbance and PL spectral profile of the sample did not change before and after washing (see Figure 6B). Thus, the decrease in the PL QY is due to the increasing number of unpassivated surface trap states. In a next step, the surface of the sample was passivated again by OAm/OTA via the addition of the ligand solution. Afterwards, we found that the QY for the passivated sample increased to 0.96% (cf. Figure 6B, orange), which is basically consistent with that of the initial sample. This ligand-related luminescence change proves the existence of numerous surface trap states in ZnS:Mn NPLs. Furthermore, it is found that the washed samples showed no other emission except the dopant emission around 590 nm (Figure S11), indicating that the energy released by the surface trap states was through a non-radiative process.

The assumption that the surface trap states resulted in a low dopant-PL QY was further confirmed by employing the following post-synthetic treatment method.[43] In this approach, a zinc-containing precursor solution ($ZnCl_2$ in hexane) was used to passivate the surface of doped NPLs. After addition of the $Zn^{2+}$ ion solution to the doped NPL dispersion (about 18 Mn ions per NPL), the QY of the sample increased substantially from 0.93% to 2.32% (cf. Figure 6C, right). The UV-Vis absorbance and dopant-PL peak positions of the sample remained nearly unaltered (cf. Figure 6C, left), suggesting that the thickness and uniformity of the NPLs were not affected significantly during this treatment. A slight blue shift (20 meV) in the maximum position of the absorption peak can be assigned to the increasing Coulomb screening (see Figure 6C, inset).[43] For comparison, we also introduced a $S^{2-}$ solution (hexane) into the NPL dispersion. It was found that the QY of the sample significantly dropped compared to the original one (cf. Figure S12). This decrease in the PL QY is a result of the quenching effect of $S^{2-}$ ions due to photogenerated hole trapping.[44-45] This post-treatment with a $Zn^{2+}$ ion solution is also applicable to other samples with different doping levels. Figure 6D shows that the dopant-PL QY of all samples (from 0.02% to 16.64%) was enhanced after the surface passivation. The enhancement factor for all samples was calculated to be approx. 2.5 (see Figure S13). This is reasonable since all samples have an identical surface-to-volume ratio.

**CONCLUSIONS**

In conclusion, we demonstrated the first direct doping of $Mn^{2+}$ ions into colloidal ultrathin ZnS NPLs via the nucleation-doping strategy. The obtained ZnS:Mn NPLs exhibit Mn luminescence via energy transfer from host to dopant. The correlation between the doping level and the evolution of PL properties as well as electron spin resonance conditions were established. The methodology allows the precise determination of the atomic environment of the dopant ions and thus, their relative location. Based on that it further allows identifying the relaxation mechanisms for the excited charge carriers in NPLs. Additionally, the surface of the NPLs was found to be a critical and limiting factor for the effective energy conversion. We show that the PL QY of ZnS:Mn NPLs can be substantially enhanced by passivating the surface of the samples with an appropriate ion solution. This opens a path for further strategies to improve the PL QY of NPLs, exemplified here by Mn doped ZnS NPLs.

**EXPERIMENTAL SECTION**

**Chemicals.** Zinc chloride (97+%), octylamine (OTA, 99+%), and methanol (99.8%) were purchased from Acros. Sulfur powder (99.998%), manganese (II) acetate {98%, $Mn(OAc)_2$} and oleylamine (OAm, 70%) were ordered from Sigma-Aldrich. Toluene (99.5%), isopropanol (99.7%), and hexane (95%) were purchased from VWR. Acetone (99%) was purchased from Th. Geyer. Chemicals. All chemicals were used without further purification.

**Synthesis of ZnS:Mn NPLs.** In a three necked flask equipped with a septum and a thermocouple in a glass mantle, 0.15 mmol (20.4435 mg) of $ZnCl_2$, 0.45 mmol (14.427 mg) of sulfur powder and a certain amount of $Mn(OAc)_2$ were dissolved in a mixture of 5 mL OAm and 10 mL OTA. The mixed solution was bubbled with nitrogen at 100 °C for 30 min under vigorous stirring. Afterwards, the reaction solution was heated to 150 °C for 6 hours with magnetically stirring under nitrogen flow. After the reaction, the solution was naturally cooled down to room temperature. The resulting nanocrystals were purified as follows: (1) 15 mL of reaction solution was mixed with 5 mL acetone and 5 mL isopropanol. (2) The mixture was shaken well and centrifuged for 10 min at 9000 rpm. (3) The supernatant was discarded and the precipitation was dispersed into 3 mL hexane or toluene for further characterization.

For the synthesis of ZnS:Mn NPLs with different Mn doping levels, the amount of $Mn(OAc)_2$ was varied by fixing the amounts of $ZnCl_2$ and sulfur powder in the initial mixture described above.

**Transmission electron microscopy (TEM).** TEM images were taken by using a JEOL Jem-1011 microscope at an acceleration voltage of 100 kV. Samples for the TEM analysis were prepared by drop-casting 10 μL of the dilute nanocrystal dispersion onto carbon-coated copper grids. High resolution (HR) TEM images and energy dispersive X-ray spectroscopy (EDS) spectra were obtained with a Philips CM 300 UT microscope operated at an acceleration voltage of 200 kV.

**X-ray diffraction (XRD).** XRD measurements were performed with a Philips X'Pert PRO MPD diffractometer with monochromatic X-Ray radiation from a copper anode with a wavelength of 0.154 nm ($CuK_α$). Samples for the XRD analysis were prepared by drop-casting a few microliter of the concentrated nanocrystal solution onto silicon wafer substrates with subsequent solvent evaporation.

**Steady-state UV-Vis absorbance and photoluminescence (PL) spectra.** Steady-state UV-Vis absorbance and PL spectra were obtained with a PerkinElmer Lambda 25 two-beam spectrometer and a Horiba Fluoromax-4 spectrometer, respectively. Time resolved PL measurements were performed with Picoquant FluoTime 300 fluorescence spectrometer. Samples were prepared by adding a few microliter of the nanocrystal solution into 3 mL of hexane in quartz vessels with an optical path length of 10 mm.

**Inductively coupled plasma optical emission spectrometry (ICP-OES).** Mn concentration in NPLs was determined via ICP-OES (type: ARCOS) produced by Fa. Spectro. Samples for the ICP-OES analysis were prepared with three steps: (1) 2 mL of the concentrated nanocrystal solution was pipetted into a 25 mL beaker and evaporated at room temperature. (2) The dried sample was dissolved in 4 mL of concentrated $HNO_3$ at 160 °C. (3) The mixed solution was diluted in a 25 mL volumetric flask.

**X-Ray Photoelectron Spectrometry (XPS).** XPS was carried out on a Kratos Axis Supra instrument using a monochromatic Al K$_\alpha$ x-ray source operated at 225 W (15 mA emission current). For each sample wide scans were carried out at a pass energy of 160 eV over the 1200 - 0 eV binding energy range to identify all the elements present on the surface. Based on the survey scans, high resolution spectroscopy was recorded for each element as a pass energy of 40 eV. Typically high resolution spectra were recorded with a step size of 0.1 eV, with a 1000 ms dwell time and repeated sweeps to improve the signal:noise. The charge neutralizer was used to eliminate any differential charging, and subsequently the binding energy axis was charge corrected to the CxHy component of the carbon components at 284.8 eV. Data was quantified using CasaXPS (2.3.23rev1.1K) using the Kratos sensitivity factor library based on Shirley backgrounds.

**Electron paramagnetic resonance (EPR) spectroscopy.** EPR measurements were recorded at 100 K on a Bruker EMX CW-micro X-band EPR spectrometer equipped with an ER4119HS high-sensitivity resonator, with a microwave power of Ca 6.9 mW, modulation frequency of 100 kHz, and amplitude of 5 G. The EPR spectrometer was equipped with a temperature controller and liquid $N_2$ cryostat for low temperature measurements. For each measurement, 2 mL of ZnS:Mn NPLs with different Mn doping levels (in hexane) was used. For calculation g values, the equation $h\nu = g\beta B_0$ was used with $\beta, B_0$ and $\nu$ being the Bohr magneton, resonance field and frequency, respectively. Calibration of the g values was performed using a DPPH standard (g = 2.0036 $\pm$ 0.0004). The EPR spectra of the samples were simulated using the software package Easyspin implemented in MATLAB.

**ASSOCIATED CONTENT**

Supporting Information

The Supporting Information is available free of charge on the ACS Publications website at http://pubs.acs.org.

Additional EDS, steady state PL and transient PL spectra, XRD patterns, XPS spectrum; Mn/Zn ratio analysis; Monte-Carlo simulation; QY per Mn analysis; PL enhancement factor (PDF)

**AUTHOR INFORMATION**

Corresponding authors

*E-mail: alf.mews@chemie.uni-hamburg.de

*E-mail: christian.klinke@uni-rostock.de

ORCID

Liwei Dai: 0000-0002-4360-6075

Christian Klinke: 0000-0001-8558-7389

Notes

The authors declare no competing financial interest.

**ACKNOWLEDEMENTS**

L.D. thanks the China Scholarship Council (CSC) for financial support. We thank Andreas Kornowski, Stefan Werner, and Almut Barck (University of Hamburg) for TEM and XRD measurements.